\def\Pbar{\overset{\rule{2mm}{.2mm}}{P}}
\def\Mbar{\overset{\rule{2mm}{.2mm}}{M}}
\def\Psibar{\overset{\rule{2mm}{.2mm}}{\Psi}}
\def\Gbar{\overset{\rule{2mm}{.2mm}}{\Gamma}}
\def\Kbar{\overset{\rule{2mm}{.2mm}}{K}}
\def\ubar{\overset{\rule{2mm}{.2mm}}{u}}
\def\ud{\mathrm{d}}

\documentclass{ws-procs9x6}

\begin{document}

\title{Electromagnetic Form Factors of Hyperons in a Relativistic
  Quark Model}

\author{ \underline{T. Van
Cauteren}\footnote{e-mail:\uppercase{T}im.\uppercase{V}an\uppercase{C}auteren@\uppercase{UG}ent.be},
T. Corthals, S. Janssen and J. Ryckebusch}

\address{Ghent University, B-9000 Gent, Belgium}

\author{D. Merten, B. Metsch and H.-R. Petry}

\address{Helmholtz-Institut f\"ur Strahlen- und Kernphysik \\
Bonn University, D-53115 Bonn, Germany}

\maketitle

\abstracts{The relativistically covariant constituent quark model
developed by the Bonn group is used to compute the EM form factors of
strange baryons. We present form-factor results for the ground-state
and some excited hyperons. The computed magnetic moments agree well
with the experimental values and the magnetic form factors follow a
dipole $Q^2$ dependence.}

\section{Motivation}\label{sec:motivation}

The photo- and electroproduction of mesons from the nucleon is a
process in which both electromagnetic and strong interactions
occur. One particularly interesting process is the electroproduction
of kaons, where a strange quark/antiquark pair is produced from the
QCD vacuum. Data from Jefferson Lab have recently been
published~\cite{mohring03}, but an appropriate theoretical description
using isobar models is still lacking. One of the main uncertainties in
these models is the incompleteness or absence of any knowledge about
the form factors, strong or electromagnetic, of the nucleon and
hyperon resonances. This work focusses on the latter. We have used the
constituent quark (CQ) model developed by the Bonn group to calculate
the electromagnetic form factors of ground-state hyperons and the
helicity amplitudes of hyperon resonances. The Bonn CQ model is
Lorentz covariant and is therefore well suited to describe baryon
properties up to high $Q^2$, which involve large recoil
effects~\cite{merten02}.

\section{EM Form Factors in the Bethe-Salpeter Approach}\label{sec:model}

\subsection{The Bethe-Salpeter Equation}\label{sec:BSE}

The Bethe-Salpeter (BS) amplitude $\chi_{\Pbar}$ is the analogue of
the wave function in the Hilbert space of three quarks with Dirac,
flavor and color degrees of freedom. Starting from the six-point
Green's function, in momentum space, the following integral equation
for the BS amplitude can be derived~:
\begin{equation}
\chi_{\Pbar} \; = \; -i \, G_{0 \Pbar} \, \left( K^{(3)}_{\Pbar} +
\Kbar^{(2)}_{\Pbar} \right) \, \chi_{\Pbar} \; .
\label{eq:BSE}
\end{equation}
This expression incorporates all features of the model. It is Lorentz
covariant in its inception, and the integral kernel is the product of
the free three-quark propagator $G_{0 \Pbar}$ and the sum of all
three- and two-particle interactions $K^{(3)}_{\Pbar} +
\Kbar^{(2)}_{\Pbar}$. The free three-quark propagator is approximated
by the direct product of three free CQ propagators. We use a linear
three-quark confinement potential for $K^{(3)}_{\Pbar}$ and the 't
Hooft instanton induced interaction for $\Kbar^{(2)}_{\Pbar}$. Both
interactions are assumed to be instantaneous.

Once the BS amplitudes are known, one can calculate any matrix element
between two baryon states, provided that the operator is known. When
computing electromagnetic form factors, the operator of interest is
the electromagnetic current operator. We use the operator $j^E_\mu =
\Psibar \hat{q} \gamma_\mu \Psi$, which describes the photon coupling
to a structureless CQ in first order of the electromagnetic
interaction. $\Psi$ and $\Psibar$ are the CQ destruction and creation
operators and $\hat{q}$ is the CQ charge operator. The current matrix
element (CME) is then computed in the c.o.m. frame of the incoming
baryon (${\Pbar}' = \Mbar$) according to~:
\begin{multline}
\langle \, \Pbar \, | \, j^\mu \, | \, \Mbar \, \rangle \simeq -3 \iint
\frac{\ud^4 \left[ \frac{1}{2} \left( p_1 - p_2 \right) \right]} {(2
\pi )^4} \frac{\ud^4 \left[ \frac{1}{3} \left( p_1 + p_2 - 2p_3\right)
\right]} {(2 \pi )^4} \\
\times \, \Gbar^{\Lambda}_{\Pbar} \; S^1_F (p_1) \otimes S^2_F (p_2) \otimes
\left[ S^3_F (p_3 + q) \, \hat{q} \gamma^\mu S^3_F (p_3) \right] \;
\Gamma^{\Lambda}_{\Mbar} \; ,
\label{eq:CME}
\end{multline}
where $\Gamma$ and $\Gbar$ are the amputated BS amplitude and its
adjoint, and $S^i_F$ is the $i$'th CQ propagator~\cite{merten02}.

\subsection{Form Factors and Helicity Amplitudes}\label{sec:FF_HA}

The electromagnetic properties of particles are usually presented in
terms of form factors, which are functions of the independent scalars
of the system. The most frequently used expression for the spin-1/2
EM-vertex is~:
\begin{multline}
\langle \, B', {\Pbar}', \lambda' \, | \, j^E_\mu (0) \, | \, B, \Pbar,
\lambda \, \rangle \, = \, e \ \ubar_{\lambda'} ({\Pbar}') \Gamma_\mu
u_\lambda (\Pbar) \\
= \, e \ \ubar_{\lambda'} ({\Pbar}') \left[ \,
\gamma_\mu F^{B'B}_1(Q^2) + \frac{i \sigma_{\mu \nu} q^\nu} {2 M_p}
\kappa_{B'B} F^{B'B}_2(Q^2) \right] u_\lambda (\Pbar) \; ,
\label{eq:EMvertex_dirac_pauli}
\end{multline}
where we have introduced the Dirac and Pauli (transition) form factors
$F^{B'B}_1$ and $F^{B'B}_2$, and the anomalous (transition) magnetic
moment $\kappa_{B'B}$. Often, the elastic form factors of the
ground-state hyperons are expressed in terms of the Sachs' electric
and magnetic form factors~:
\begin{eqnarray}
G^B_E(Q^2) = F^B_1(Q^2) - \frac{Q^2} {4 M_B M_p} \kappa_B
F^B_2(Q^2) = \frac{\langle B, {\Pbar}', \frac{1}{2} | j^E_0 (0) | B,
\Pbar, \frac{1}{2} \rangle } {\sqrt{4M^2_B + Q^2}} \;
\label{eq:sach_el} \\
G^B_M(Q^2) = F^B_1(Q^2) + \frac{M_p} {M_B} \kappa_B F^B_2(Q^2) =
\frac{\langle B, {\Pbar}', \frac{1}{2} | j^E_+ (0) | B, \Pbar,
-\frac{1}{2} \rangle} {2 \sqrt{Q^2}} \; . \; \label{eq:sach_ma}
\end{eqnarray}

The response of hyperon resonances is commonly expressed in terms of
helicity amplitudes, which are directly proportional to the spin-flip
($A_{1/2}$ and $A_{3/2}$) and non-spin-flip ($C_{1/2}$) CME's, with
proportionality constant $\sqrt{\frac{\pi \alpha} {2 M^* ({M^*}^2 -
M^2)}}$.

\section{Results and Conclusions}\label{sec:results}

\begin{table}[ph]
\vspace*{-7pt}
\tbl{Static electromagnetic properties of the ground-state
hyperons. Magnetic moments are given in units of $\mu_N$, square radii
in units of fm$^2$.}
{\footnotesize
\begin{tabular}{@{}ccccc@{}}
\hline
{} \hspace{2cm} &{} \hspace{1.5cm} &{} \hspace{1.5cm} &{} \hspace{1.5cm} &{}
\hspace{1.5cm}\\[-2.5ex]
Y & $\mu^{\text{exp}}_Y$ & $\mu^{\text{calc}}_Y$ &
$<r^2_M>$ & $<r^2_E>$\\[0.4ex]
\hline
{} \hspace{2cm} &{} \hspace{1.5cm} &{} \hspace{1.5cm} &{} \hspace{1.5cm} &{}
\hspace{1.5cm}\\[-2.5ex]
$\Lambda$ &-0.613 &-0.61 &0.40 &0.038\\
$\Sigma^+$ &2.458 &2.47 &0.69 &0.79\\
$\Sigma^0$ &----- &0.73 &0.69 &0.150\\
$\Sigma^-$ &-1.160 &-0.99 &0.81 &0.49\\
$|\Sigma^0 \rightarrow \Lambda |$ &1.61 &1.52 &1.96 &-0.120\\
$\Xi^0$ &-1.25 &-1.33 &0.47 &0.140\\
$\Xi^-$ &-0.65 &-0.57 &0.38 &0.47\\
\hline
\end{tabular}\label{tab:gs_hyp} }
\vspace*{-13pt}
\end{table}

In Tables~\ref{tab:gs_hyp} and~\ref{tab:res_hyp}, we summarize the
obtained results for the static properties of the ground-state hyperons and
resonances respectively. The magnetic moments are generally in very
good agreement with the experimental values. The electric mean-square
radius of the $\Sigma^-$ is in agreement with the values of $0.91
\pm 0.32 \pm 0.40$~fm$^2$ of Adamovich \emph{et
al.}~\cite{adamovich99} and $0.60 \pm 0.08 \pm 0.08$~fm$^2$ from
Eschrich \emph{et al.}~\cite{eschrich01}. The decay widths of the
hyperon resonances are poorly known. However, from
Table~\ref{tab:res_hyp} it is clear that the one for the
$\Lambda$(1405) is badly reproduced in our model, which may indicate
the special nature of this resonance~\cite{hyodo04}.

\begin{table}[ph]
\vspace*{-7pt}
\tbl{Static electromagnetic properties of the hyperon resonances for
which experimental results are available. Photo-amplitudes are given
in units of 10$^{-3}$ GeV$^{-1/2}$ and widths in units of MeV.}
{\footnotesize
\begin{tabular}{@{}ccccc@{}}
\hline
{} \hspace{2cm} &{} \hspace{1.3cm} &{} \hspace{1.3cm} &{} \hspace{1.3cm} &{}
\hspace{1.3cm}\\[-2.5ex]
$Y^*$ & $|A_{1/2}|$ & $|A_{3/2}|$ & $\Gamma_{\text{calc}}$ &
$\Gamma_{\text{exp}}$\\[0.4ex]
\hline
{} \hspace{2cm} &{} \hspace{1.3cm} &{} \hspace{1.3cm} &{} \hspace{1.3cm} &{}
\hspace{1.3cm}\\[-2.5ex]
P$_{13}$(1385) &62.8 &108 &1.46 &0 -- 13.9\\
S$_{01}$(1405) &51.5 &----- &0.912 & 0.019 -- 0.035\\
D$_{03}$(1520) &5.50 &41.2 &0.258 &0.0876 -- 0.166\\
\hline
\end{tabular}\label{tab:res_hyp} }
\vspace*{-13pt}
\end{table}

The elastic Sachs' electric and magnetic form factors of the
ground-state hyperons, as well as the transition Dirac and Pauli form
factors of the $\Sigma^0 \rightarrow \Lambda$ transition, are presented
in Ref.~\cite{tvc04}. There it is shown that the
magnetic form factors have a dipole dependence on $Q^2$ with cutoffs
ranging from 0.79 to 1.14 GeV. We also observed that some electric
form factors change sign at a finite value of $Q^2$.

\begin{figure}[ht]
\vspace*{-5pt}
\centerline{\epsfxsize=4.1in\epsfbox{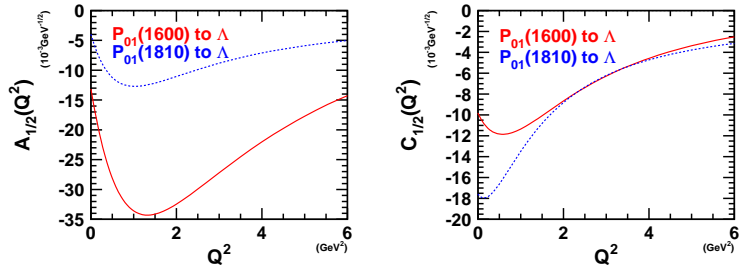}}   
\caption{Helicity amplitudes $A_{1/2}$ and $C_{1/2}$ for the
P$_{01}$(1600) (full line) and the P$_{01}$(1810) (dashed line)
resonances. \label{fig:p01_hyp}}
\vspace*{-24pt}
\end{figure}

The results for the helicity amplitudes of two P$^*_{01}$ to $\Lambda$
decays are displayed in Fig.~\ref{fig:p01_hyp}. For the P$_{01}$(1600),
the $A_{1/2}$ peaks at a finite value of $Q^2$. Accordingly, our
results indicate that resonances which are of minor importance in
photoproduction reactions can play a major role in the corresponding
electroproduction process.

 \end{document}